\documentclass[a4paper]{mem}
\usepackage{natbib}
\usepackage{graphicx}
\usepackage[a4paper]{hyperref}
\begin{document}
\title{Future Japanese missions for the study of warm-hot
intergalactic medium }

\author{T. Ohashi\inst{1}, M. Ishida\inst{1}, S. Sasaki\inst{1},
   Y. Ishisaki\inst{1}, K. Mitsuda\inst{2}, \\ N. Y. Yamasaki\inst{2},
   R. Fujimoto\inst{2}, T. Furusho\inst{2}, H. Kunieda\inst{2}, \\
   Y. Tawara\inst{3}, A. Furuzawa\inst{3}, Y. Suto\inst{4}, \and
   K. Yoshikawa\inst{4}}

\offprints{T. Ohashi} \mail{Hachioji, Tokyo 192-0397, Japan}

\institute{Department of Physics, Tokyo Metropolitan University, 1-1
Minami-Ohsawa, Hachioji, Tokyo 192-0397, Japan \\
\email{ohashi@phys.metro-u.ac.jp}\\
\and 
Institute of Space and Astronautical Science, Japan Aerospace
Exploration Agency, 3-1-1 Yoshinodai, Sagamihara, Kanagawa 229-8510,
Japan \\
\and 
Department of Astrophysics, Nagoya University, Furo-cho, Chikusa,
Nagoya 464-8602, Japan \\
\and 
Department of Physics, University of Tokyo, 7-3-1 Hongo, Bunkyo-ku,
Tokyo 113-0033, Japan}

\abstract{ We present our proposal for a small X-ray mission DIOS
(Diffuse Intergalactic Oxygen Surveyor) to perform survey
observations of warm-hot intergalactic medium using OVII and OVIII
emission lines.  This will be proposed to a small satellite program
planned by ISAS/JAXA in Japan for a launch in 2008. The instrument
consists of an array of TES microcalorimeters with an energy
resolution 2 eV, cooled by mechanical coolers. The X-ray telescope
will employ 4-stage reflection mirrors with a focal length as short as
70 cm and an angular resolution $2'$.  In addition to DIOS, we briefly
describe the NeXT (New X-ray Telescope) mission, which is a larger
Japanese X-ray observatory to be launched in 2010 and plans to explore
non-thermal processes in the universe.

 }

\authorrunning{T. Ohashi et al.}
\titlerunning{Future Japanese missions} \maketitle
%

\section{Introduction}
Following the X-ray spectroscopic mission Astro-E2 to be launched in
2005 \citep[see][]{furusho04,ohashi01}, Japanese X-ray astronomy groups are
considering several satellite missions in the future. Here, we will
describe a dedicated small mission to explore the structure of diffuse
intergalactic medium, called DIOS (Diffuse Intergalactic Oxygen
Surveyor), to be launched in 2008. We also briefly describe a larger
X-ray mission NeXT (New X-ray Telescope mission) to appear in
2010. These missions will bring a substantial advance in our
understanding of the hot-gas distribution in the universe, as well as
in our own galaxy.

\section{DIOS}
DIOS is proposed to fit in a small satellite program which is a new
scheme under consideration by ISAS/JAXA\@. The main purpose of the
mission is a sky survey of warm-hot intergalactic medium using oxygen
K emission lines, made possible by superior energy resolution and
large field of view. The combination of a 4-stage X-ray telescope and
a large array of TES microcalorimeter are the new technologies
involved. See \citet{yoshikawa03} for a detailed description of the
science carried out from DIOS including simulated spectra. This
mission can also perform a mapping observation of the hot interstellar
medium in our galaxy. The high energy resolution ($\Delta E \approx 2$
eV) will reveal the Doppler shifts of the hot interstellar gas with a
velocity $\sim 100$ km s$^{-1}$. The observations from DIOS can
confirm the large-scale outflow and falling-in of the hot bubbles
in our galaxy (galactic fountain).
\begin{figure}[!] \begin{center}
\includegraphics[width=0.47\textwidth]{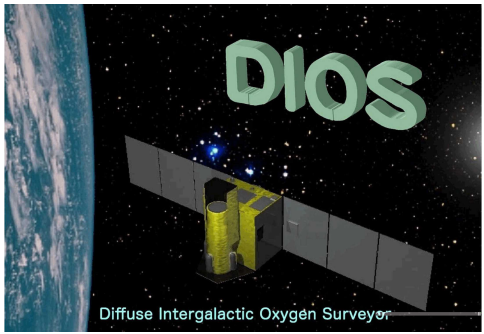}
\caption{The DIOS spacecraft. The length of the solar paddle is 6 m.}
\end{center}\end{figure}

\subsection{Spacecraft}

The view of the DIOS spacecraft is shown in Fig.\ 1 and main
parameters of the satellite are listed in Table 1. The spacecraft will
weigh about 400 kg, out of which the payload takes $\sim 280$ kg. This
mass enables the satellite to be launched as a piggy-back or
sub-payload in H2 or Ariane rockets. It is also possible for a launch
with the new ISAS rocket, such as M-V light. The size before the
launch is $1.5 \times 1.5 \times 1.2$ m, and the 1.2 m side will be
expanded to about 6 m after the paddle deployment.
\begin{table}\begin{center}
\caption{Parameters of the observing instruments on board DIOS}
\begin{tabular}{|l|l|}\hline
Effective area & $> 100$ cm$^2$ \\
Field of view & $50'$ diameter \\
$S\Omega$ & $\sim 100$ cm$^2$ deg$^2$ \\
Angular resolution & $3'\ (16\times 16$ pixels) \\
Energy resolution & 2 eV (FWHM) \\
Energy range & 0.1-- 1 keV \\
Observing life & $> 5$ yr \\ \hline
\end{tabular} \end{center}
\end{table}

The total power is 500 W, of which 300 W is consumed by the
payload. The nominal orbit is a near earth circular one with an
altitude of 550 km. This low-earth orbit can be reached by the ISAS
rocket M-V light.  An alternative choice of the orbit under
consideration is an eccentric geostationary transfer orbit. This orbit
gives a lower heat input from the earth and relaxes the thermal design
of the satellite, and would enhance the launch opportunity to be
carried as a sub-payload for geostationary satellites. One significant
drawback is the increased particle background level as already met by
Chandra and XMM-Newton. In the soft energy range below 1 keV,
electrons can be a major source of background.

The attitude will be 3-axis stabilized with momentum wheels. Typical
pointing accuracy will be about $10''$. The direction of the field of
view can be varied within $90^\circ \pm 20^\circ$ from the sun
direction. With this constraint, any position in the sky can be
accessed within half a year.

\begin{figure*}[!tb] \begin{minipage}{0.53\textwidth}
\includegraphics[width=0.95\textwidth]{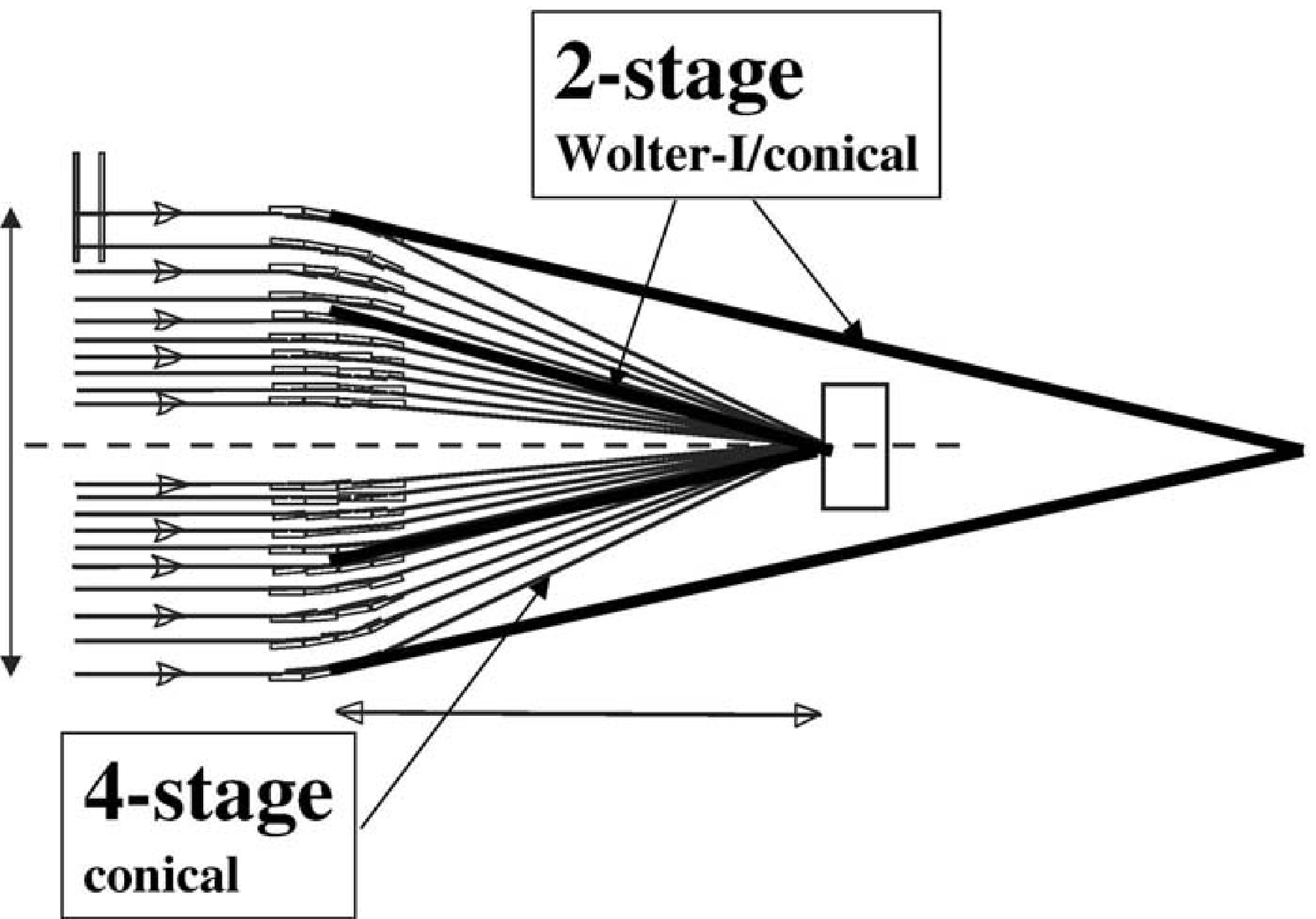}
\caption{Concept of the 4-stage reflection telescope.}
\end{minipage}\hfill
\begin{minipage}{0.43\textwidth}
\includegraphics[width=0.75\textwidth,angle=270]{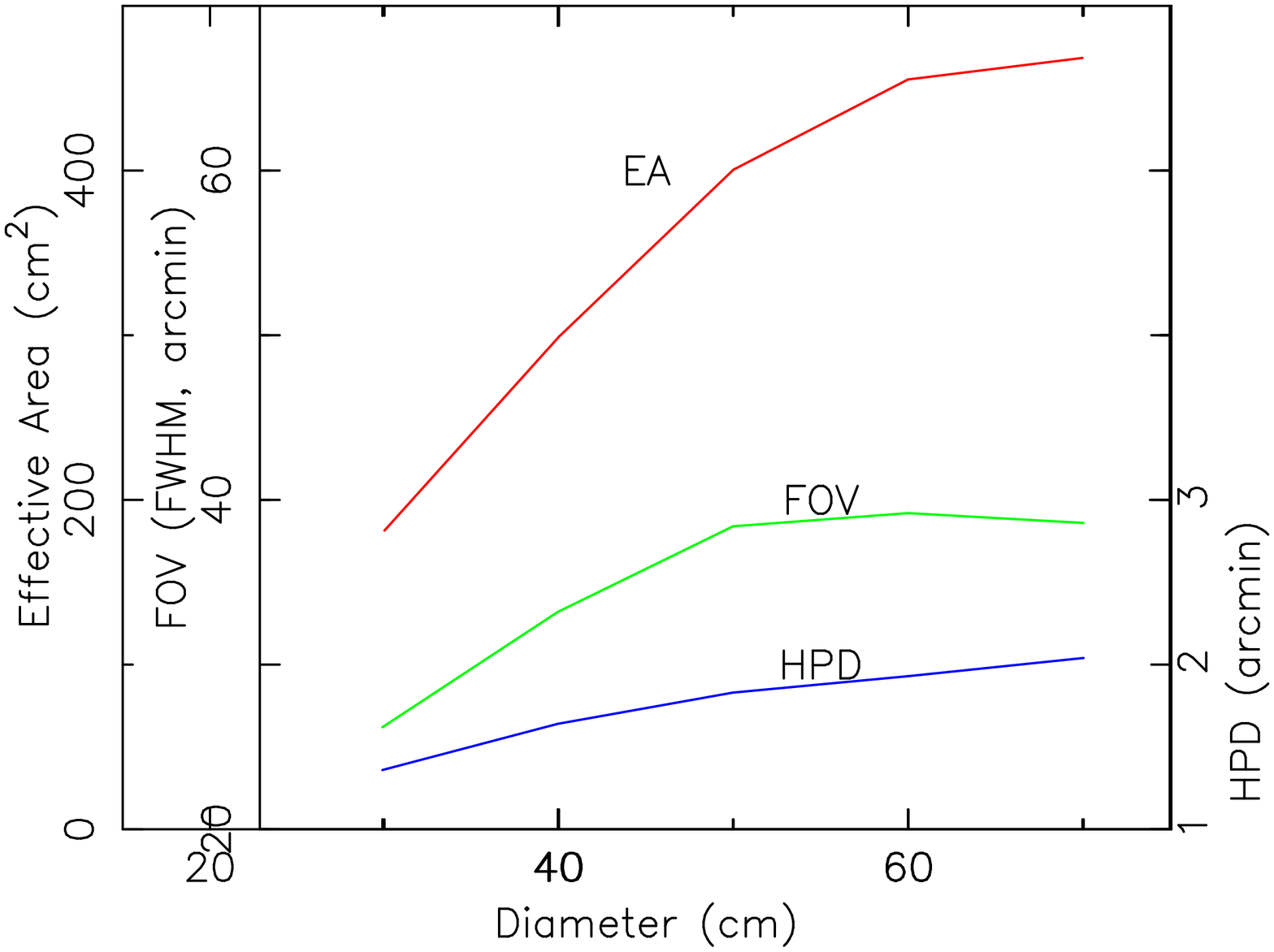}
\caption{Effective area, field of view, and half-power diameter as a function of the mirror diameter.}
\end{minipage}
\end{figure*}

\subsection{Instruments}

Several new technologies will be introduced in the DIOS mission. The
4-stage X-ray telescope FXT (Four-stage X-ray Telescope) is the first
one \citep{tawara03}. As shown in Fig.\ 2, incident X-rays are
reflected 4 times by thin-foil mirrors and are focused at only 70 cm
from the mirror level. At the energy of oxygen lines, $\sim 0.6$ keV,
the reflectivity of the mirror surface is as high as 80\% and the
reduction of the effective area is not a serious problem. The 4-stage
reflection gives the focal length half as long as the usual 2-stage
design, and substantially saves the volume and weight of the
satellite. Also, a relatively small focal plane detector can have a
wide field of view, which is a great advantage for a TES calorimeter
array. In our basic design, the outer diameter of the mirror and the
effective area are 50 cm and 400 cm$^2$, respectively, at 0.6 keV as
shown in Fig.\ 3. Ray-tracing simulation indicates that the angular
resolution of is 2 arcmin (half-power diameter), and the image quality
does no show significant degradation at an offset angle $30'$. This
4-stage telescope is a key design factor in making DIOS accommodated in
the small satellite package.

The focal plane instrument XSA (X-ray Spectrometer Array) is an array
of TES microcalorimeters, whose development in Japan is a
collaboration with Waseda University, Seiko Instruments Inc., and
Mitsubishi Heavy Industries. There will be $16\times 16$ pixels
covering an area of about 1 cm square. The corresponding field of view
is $50'$. XSA will have an energy resolution of 2 eV FWHM at 0.6
keV\@. An example of pulse-height spectra obtained for 5.9 keV X-rays
with our TES calorimeter, which is a single pixel type made with Ti-Au
bilayer with Au X-ray absorber, is shown in Fig.\ 4. The measured
resolution is 6.3 eV FWHM, so there still needs some further
development, as well as a tuning of the calorimeter design, to achieve
the best performance for 0.6 keV X-rays.  We are now developing
several new techniques toward the multi-pixel operation of TES
calorimeters \citep(ishisaki03,iyomoto03). X-ray absorbing material
under consideration is Bi, which has low heat capacity and does not
produce long-living quasi particles in the absorber. We are testing an
electro-plating method to build a $16 \times 16$ array with a pixel
size $\sim 0.5 \times 0.5$ mm supported by a thin stem (see Fig.\ 5
for our test model made with Sn). For the signal readout, efficient
multiplexing of the signals is essential to take all the data out from
the cold stage. We are trying to add the signals in frequency space by
operating the TES calorimeters with AC bias at different
frequencies. We could successfully decode signals from 2 different
pixels so far.  A new multi-input SQUID has also been developed to add
the signals from several TES pixels together. An efficient thermal
shield with high soft X-ray transmission is an essential item, and we
are looking at very thin Be foils for this purpose.
\begin{figure*}[!tb] \begin{minipage}{0.56\textwidth}
\includegraphics[width=1.02\textwidth]{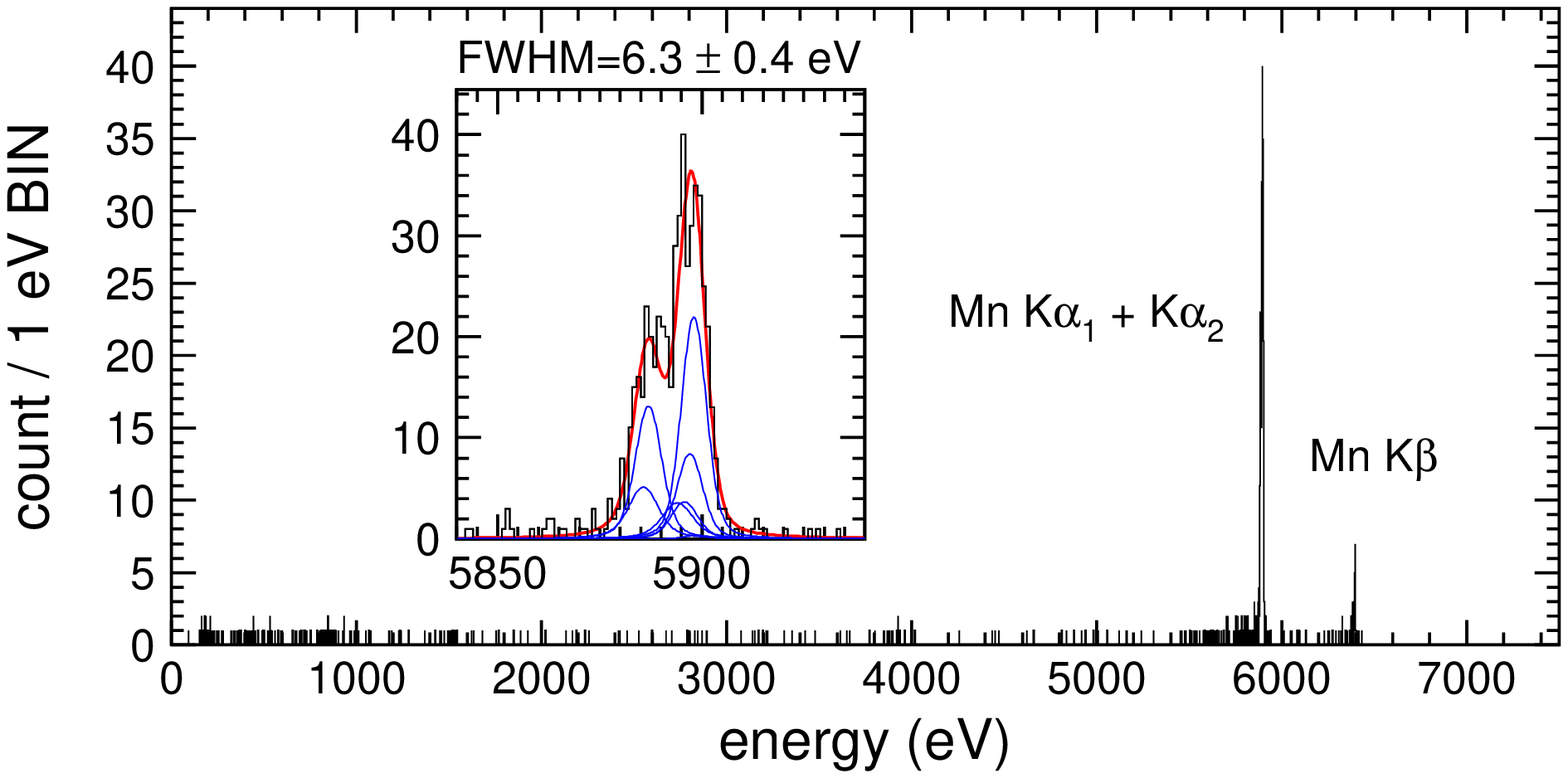}
\caption{Pulse-height spectrum for Mn-K$\alpha$ X-rays obtained by a single pixel Ti-Au TES microcalorimeter.}
\end{minipage}\hfill
\begin{minipage}{0.4\textwidth}
\includegraphics[width=0.95\textwidth]{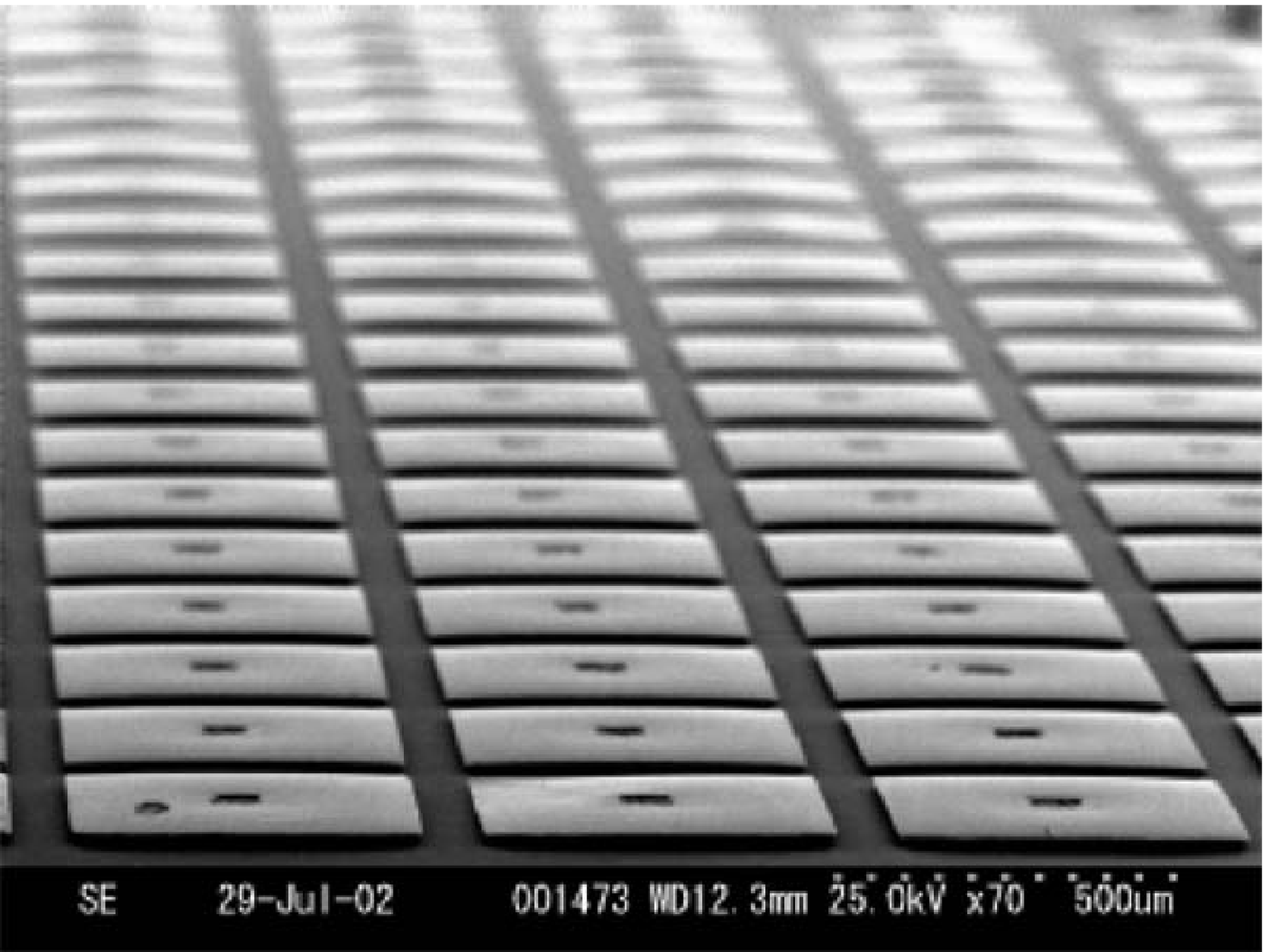}
\caption{Mechanical model of the 256 pixel absorber for the TES array, produced by a Sn plating. Size of one pixel is 500 $\mu$m.}
\end{minipage}\end{figure*}

Another important feature of DIOS is the cryogen-free cooling
system. We are considering a serial connection of different types of
coolers to achieve $\sim 50$ mK for XSA within the available power
budget. In the first stage, a Stirling cooler takes the temperature
down to 20 K, and then $^3$He Joule-Thomson cooler reduces it to 1.8 K
as the second stage. The third stage is $^3$He sorption cooler
achieving 0.4 K, and finally adiabatic demagnetization refrigerator
obtains the operating temperature at 50 mK\@. Since no cryogen is
involved, this cooling system ensures an unlimited observing life in
the orbit, which is a big advantage of DIOS\@. The XSA system is
subject to a warm launch, therefore we have to allow for the initial
cooling of the system for the first 3 months in the orbit.

Since several challenging technologies are involved, we will naturally
seek for international collaboration in various parts of the observing
instruments in due course of time.

\subsection{Observation}
As shown by the simulations in \citet{yoshikawa03}, a single pointing
will need typically 100 ksec to obtain enough oxygen photons from the
IGM\@. Fig.\ 6 shows the expected pulse-height spectrum of a 0.2 keV
plasma in the case 1000 photons of oxygen lines are collected. We
expect that this quality of data can be obtained from surrounding
regions of clusters of galaxies with 100--300 ksec of observation. For
large-scale filaments, roughly 10 times less photos are expected for
the same observing time. On the other hand, the hot interstellar
medium in our galaxy produces some 10 times stronger line
emission. With this data quality, the lines in the OVII triplet are
clearly detected and we will be able to measure the temperature
directly. These 3 lines can also be used to separate individual plasma
components when several emission regions with different redshifts are
overlapping in the same line of sight.

 \begin{figure}[!] \begin{center}
 \includegraphics[width=0.35\textwidth,angle=270]{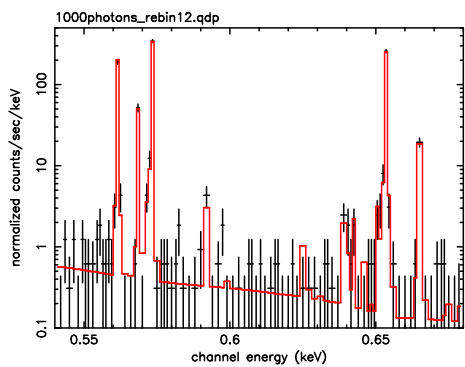}
 \caption{Pulse-height spectrum for OVII and OVIII lines expected with
 DIOS from a plasma with $kT = 0.2$ keV when 1000 line photons are
 accumulated.} \end{center}\end{figure}

Effective observing time will be approximately 40 ksec for a near
earth orbit. So if we spend 100 ksec in each pointing position, a sky
map for an area of $10^\circ \times 10^\circ$ can be produced in about
a year. This angular size is enough to see the large-scale structure
of the universe at $z < 0.3$. This survey observation of a limited sky
will be the first task of DIOS\@. Since oxygen lines from the galactic
ISM are stronger by roughly 2 orders of magnitude, 1 ksec in each point
is good to make a large map of the ISM distribution. We plan to devote
the second year for a survey of a $100^\circ\times 100^\circ$ sky for
the galactic ISM\@. After these 2 years, further deeper observations
of the IGM as well as of the outflowing hot gas in near-by galaxies
can be performed. Fig.\ 7 compares $S\Omega$ and energy resolution
among different instruments in the X-ray missions which are planned or
already operating. Clearly, DIOS will achieve the highest sensitivity
for soft X-ray lines from extended objects. We believe that DIOS, with
its complementary performance to large X-ray missions, will bring a
very rich science about the structure and evolution of the warm-hot
gas contained in galaxies and in intergalactic space.
\begin{figure} \begin{center}
\includegraphics[width=0.45\textwidth]{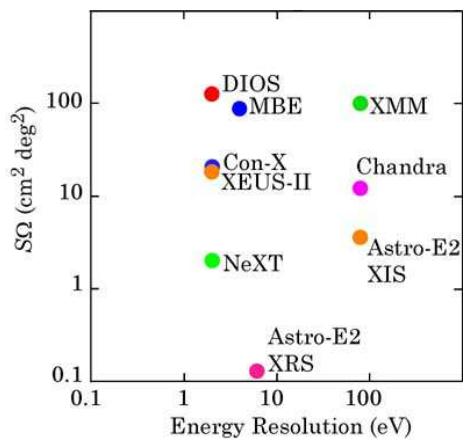}
\caption{Comparison of $S\Omega$ and energy resolution for
spectroscopic instruments (CCDs and microcalorimeters) for planned and
operating X-ray satellites.}
\end{center}\end{figure}

\section{NeXT}

NeXT (Fig.\ 8) is a larger X-ray mission with the weight ($\sim 1700$
kg) similar to that of Astro-E2, and is planned for launch in early
2010 with the ISAS M-V rocket. The major technical breakthrough
involved in this mission is the supermirror multi-layer coating on
X-ray telescopes.  This technique makes the telescope sensitive up to
80 keV, and the first true image with unprecedented sensitivity in the
hard X-ray range will be obtained. This mission will, therefore, show
us images of the regions where non-thermal processes dominate, such as
merger shocks in clusters of galaxies, shocks in supernova remnants,
and cosmic jets in active galaxies, for the first time. Compared with
thermal emission, our understanding of non-thermal processes in the
universe, such as the origin of high energy cosmic rays and the
acceleration mechanism for cosmic jets, has been very limited. We hope
that the advent of NeXT will bring us a clear jump in the observational
study of hard X-rays, just like what the Einstein observatory has
done in the soft X-ray band.
\begin{figure*}[!] \begin{minipage}{0.4\textwidth}
\includegraphics[width=0.95\textwidth]{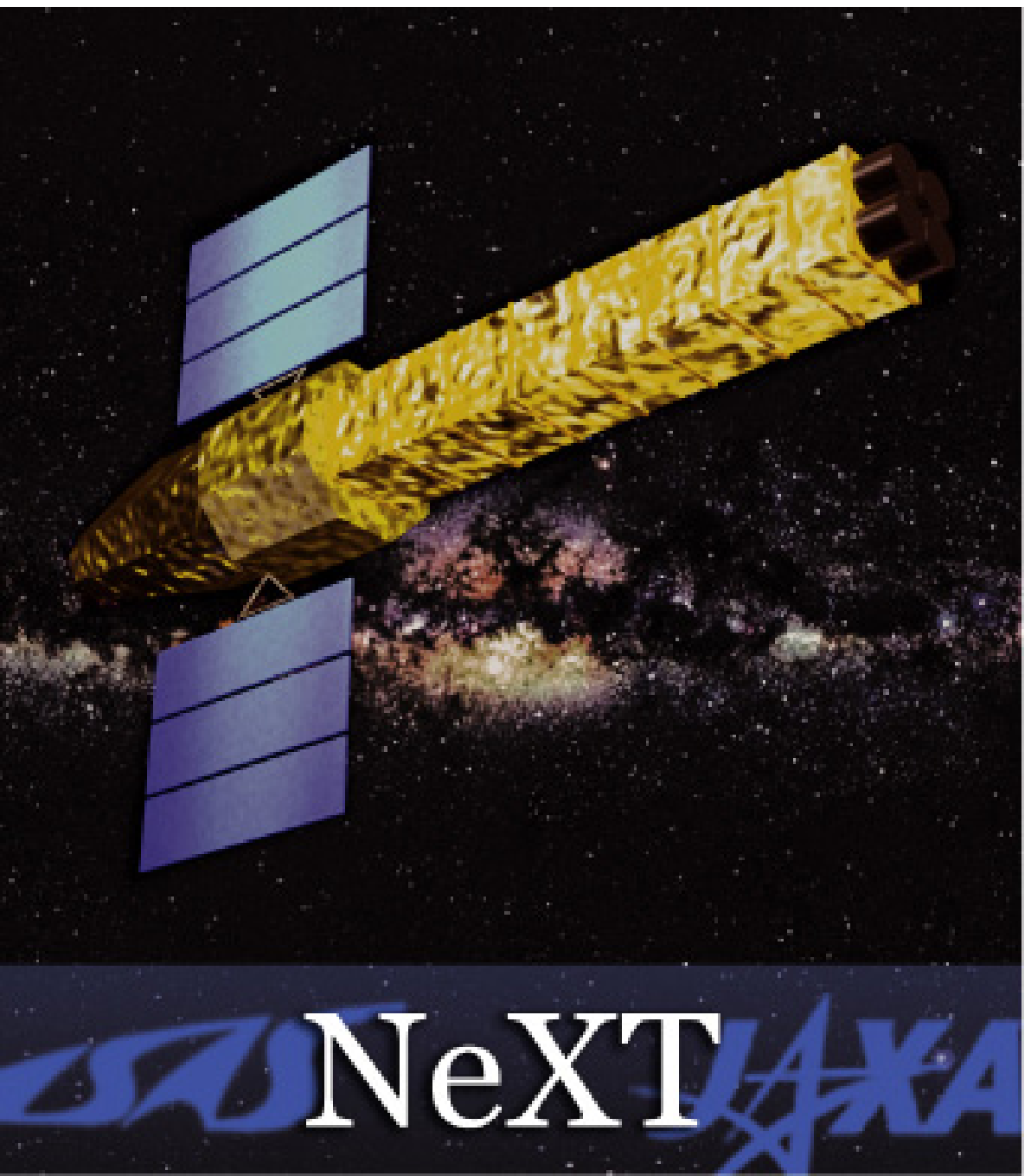}
\caption{The view of the NeXT spacecraft.}
\end{minipage}\hfill
\begin{minipage}{0.55\textwidth}
\includegraphics[width=1.0\textwidth]{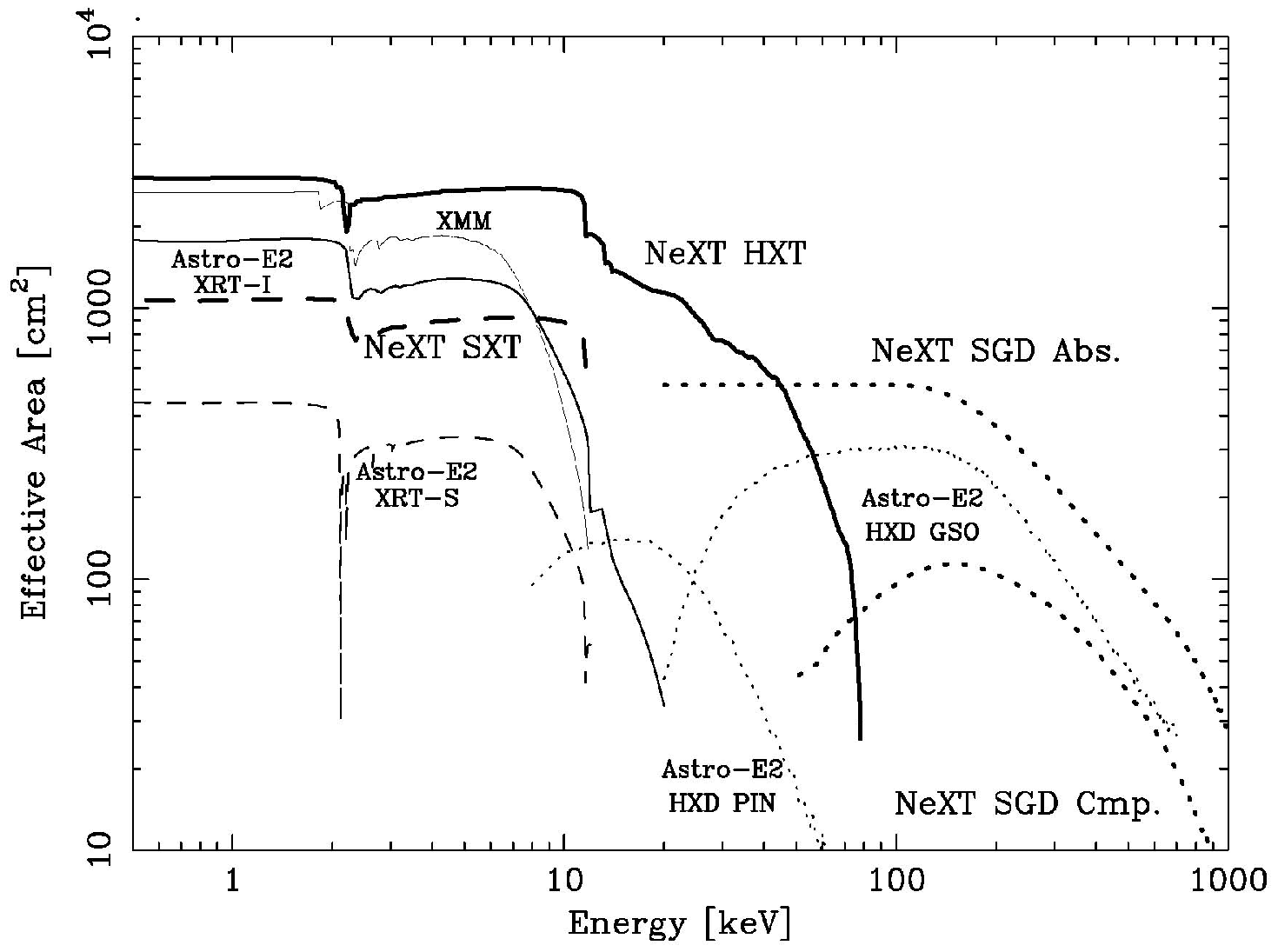}
\caption{Effective area of the experiments on board NeXT as a function
of energy, compared with those of Astro-E2.}
\end{minipage}\end{figure*}

The payload of NeXT consists of 3 main experiments, and their
effective areas against energy are shown in Fig.\ 9. The first
experiment includes 3 identical sets of the hard X-ray telescope, with
supermirror coatings, combined with hybrid imaging detectors. The
telescope system has a focal length of 12 m and gives a total area
about 700 cm$^2$ at 30 keV with angular resolution better than
$1'$. The focal plane detector comprises a Cd-Te pixel detector and an
X-ray CCD, stacked together. The hard X-rays above 15 keV penetrate
CCD chips and hit the Cd-Te detector underneath with a pixel size
$250\mu$m square.  The second experiment is an array of TES
microcalorimeters combined with one unit of soft X-ray telescope,
covering an energy range between 0.5 and 10 keV\@. Compared with
Astro-E2 XRS, the number of pixels increases from 32 to 1024, with the
effective area larger by a factor of 2.5, and the field size 4 times
larger at $6' \times 6'$. Energy resolution will also be improved to 2
eV FWHM\@. The use of mechanical coolers without cryogen enhances the
observing life in the orbit to more than 5 yrs.  The third experiment
on board is a non imaging soft $\gamma$-ray detector. This is a narrow
field Compton telescope with an active collimator and multi-layer
semiconductor detectors with an effective area about 500 cm$2$ and an
energy range up to 1 MeV\@. The well-type collimator limits the field
of view to $4^\circ$, and a requirement to fulfill the Compton
kinematics gives a very low background. Polarization measurements are
also possible above 70 keV\@.

With these experiments, we will obtain combined data on detailed gas
motion, hard X-ray image, and $\gamma$-ray spectrum from various
X-ray objects with the highest sensitivity ever achieved.
We hope this mission will lead us to the true understanding of the most
energetic processes occurring in the universe.

\bibliographystyle{aa}

\end{document}